\documentclass{article}
\usepackage{amsmath}
\usepackage{amssymb}
\usepackage{amsthm}
\usepackage{graphicx}
\usepackage{verbatim}
\usepackage{subfigure}
\usepackage{isit2004}  

\newtheorem{theorem}{Theorem}
\newtheorem{cor}{Corollary}

\newtheorem{lemma}{Lemma}

\newtheorem{example}{Example}

\begin{document}

\newcommand{\draw}[4]{
    \begin{figure}
    \begin{center}
    \includegraphics[scale=#1]{#2.eps}
    \caption{{\footnotesize #3}} \label{#4}
    \end{center}
    \end{figure}
}

\newcommand{\drawtwo}[9]{
\begin{figure}
\centering
\subfigure[{\footnotesize #3}]{     \label{#4}
\includegraphics[scale=#1]{#2.eps}
}
\subfigure[{\footnotesize #6}]{     \label{#7}
\includegraphics[scale=#1]{#5.eps}
} \caption{{\footnotesize #8}}  \label{#9}
\end{figure}
}

\newcommand{\yset}[2]{
\left\{\left(#1,y\right):#2\right\} }
\newcommand{\xset}[2]{
\left\{\left(x,#1\right):#2\right\} }
\newcommand{\xyset}[1]{\yset{x}{#1}}
\newcommand{\xysetc}[1]{\xyset{#1} & \cup \\}
\newcommand{\floor}[2]{\left\lfloor\frac{#1}{#2}\right\rfloor}
\newcommand{\ceil}[2]{\left\lceil\frac{#1}{#2}\right\rceil}

\newcommand{\field}[1]{\mathbb{#1}}
\newcommand{\C}{\field{C}}
\newcommand{\F}{\field{F}}

\newcommand{\U}{{\cal U}}
\newcommand{\Po}{{\cal P}}
\newcommand{\Sp}{{\cal S}}
\newcommand{\T}{{\cal T}}
\newcommand{\N}{{\cal N}}
\newcommand{\A}{{\cal A}}
\newcommand{\Q}{{\cal Q}}
\newcommand{\cC}{{\cal C}}
\newcommand{\Z}{\mathbb{Z}}
\newcommand{\I}{{\cal I}}
\newcommand{\G}{{\cal G}}
\newcommand{\D}{{\cal D}}
\newcommand{\R}{{\cal R}}
\newcommand{\Tr}{T_r}
\newcommand{\Tl}{T_l}
\newcommand{\sv}{\overrightarrow{s}}

\newcommand{\al}{\alpha}
\newcommand{\be}{\beta}
\newcommand{\ga}{\gamma}
\newcommand{\de}{\delta}

\newcommand{\ce}{\varsigma}
\newcommand{\Sc}[1]{\Sp(\ce_{#1})}
\newcommand{\Uc}[1]{\U(\ce_{#1})}
\newcommand{\Ur}[1]{\U_r(\ce_{#1})}
\newcommand{\Ul}[1]{\U_l(\ce_{#1})}
\newcommand{\Pc}[2]{\Po(\ce_{#1},\ce_{#2})}
\newcommand{\Nc}[1]{\N(\ce_{#1})}
\newcommand{\Nd}{\N(\ce_0,\ce_1)}
\newcommand{\Nt}{\N(\ce_1,\ce_2)}
\newcommand{\TR}[2]{TR(\ce_{#1},\ce_{#2})}
\newcommand{\Qc}{\Q(\ce)}
\newcommand{\Tc}{\T(\ce)}
\newcommand{\SNc}{SN(\ce)}
\newcommand{\Ic}[2]{\I(\ce_{#1},\ce_{#2})}
\newcommand{\rc}[1]{\Tr(\ce_{#1})}
\newcommand{\lc}[1]{\Tl(\ce_{#1})}
\newcommand{\dc}[3]{d_3(\ce_{#1},\ce_{#2},\ce_{#3})}
\newcommand{\dz}{d_3(z_1,z_2,z_3)}
\newcommand{\Fc}[1]{F(\ce_{#1})}

\topmargin = -5mm


\itwtitle{Construction of Error-Correcting Codes for Random
Network Coding}

\itwauthor{Tuvi Etzion}
{Department of Computer Science\\
Technion, Haifa 32000, Israel\\
e-mail: {\tt etzion@cs.technion.ac.il } \\
phone: 04-8294311, fax: 04-8293900 }

\itwsecondauthor{Natalia Silberstein}
           {Department of Computer Science \\
            Technion, Haifa 32000, Israel\\
            e-mail: {\tt natalys@cs.technion.ac.il} \\
            phone: 04-8294952, fax: 04-8293900
           }

\itwmaketitle

\begin{itwabstract}
In this work we present error-correcting codes for random network
coding based on rank-metric codes, Ferrers diagrams, and
puncturing. For most parameters, the constructed codes are larger
than all previously known codes.

\end{itwabstract}

\noindent Classification: Information Theory and Coding Theory

\begin{itwpaper}

\itwsection{Introduction}

The {\it projective space} of order $n$ over finite field
$\F_q=GF(q)$, denoted by $\mathcal{P}_{q}(n),$ is the set of all
subspaces of the vector space $\F_q^n$. $\mathcal{P}_{q}(n)$ is a
metric space with the distance function
$d_{S}(U,V)=\mbox{dim}(U)+\mbox{dim}(V)-2\mbox{dim}(U\cap V)$, for
all $U,V\in\mathcal{P}_{q}(n)$. A code in the projective space is a
subset of $\mathcal{P}_{q}(n)$. Koetter and Kschischang
\cite{Koetter-Kschischang} showed that codes in $\mathcal{P}_{q}(n)$
are useful for correcting errors and erasures in random network
coding. If the dimension of each codeword is a given integer $k\leq
n$ then the code forms a subset of a the Grassmannian
$\mathcal{G}_{q}(n,k)$ and called a {\it constant-dimension code}.

The {\it rank distance} between $X,$$Y\in \F_q^{m \times t}$ is
defined by $d_{R}(X,Y)=\mbox{rank}(X-Y)$. It is well
known~\cite{Gabidulin} that the rank distance is a metric. A code
$C\subseteq F_q^{m \times t}$ with the rank distance is called a
{\it rank-metric code}. The connection between the rank-metric codes
and codes in $\mathcal{P}_{q}(n)$ was explored
in~\cite{optimak_rank,Koetter-Kschischang,Silva}.

We represent a $k$-dimensional subspace $U\in\mathcal{P}_{q}(n)$ by
a $k\times n$ matrix, in {\it reduced row echelon form}, whose rows
form a basis for $U$. The\emph{ echelon Ferrers form} of a binary
vector $v$ of length $n$ and weight $k$, $EF(v)$, is a $k\times n$
matrix in reduced row echelon form with leading entries (of rows) in
the columns indexed by the nonzero entries of $v$ and $"\bullet"$
(will be called {\it dot}) in the ``arbitrary'' entries.
\begin{example} Let $v=0110100$. Then \[
EF(v)=\left(\begin{array}{ccccccc}
0 & 1 & 0 & \bullet & 0 & \bullet & \bullet\\
0 & 0 & 1 & \bullet & 0 & \bullet & \bullet\\
0 & 0 & 0 & 0 & 1 & \bullet & \bullet\end{array}\right).\]
\end{example}

Let $S$ be the sub-matrix of $EF(v)$ that consists of all its
columns with dots. A matrix $M$ over $\F_q$ is said to be in
$EF(v)$ if $M$ has the same size as $S$ and if $S_{i,j}=0$ implies
that $M_{i,j}=0$. $EF(v[M])$ will be the matrix that result by
placing the matrix $M$ instead of $S$ in $EF(v)$.
\itwsection{Construction of Constant Dimension Codes} Let $\cC$ be
a constant-weight code of length $n$, constant weight $k$, and
minimum Hamming distance $d_{H}=2\delta$. Let $C_{v}$ be the
largest rank-metric code with the minimum distance $d_{R}=\delta$,
such that all its codewords are in $EF(v)$. Now define code
$\C=\underset{{\scriptstyle v\in \cC}}{\cup}\left\{ EF(v[c]):\;
c\in C_{v}\right\}$.

\begin{lemma}
For all $v_{1},v_{2}\in \cC$ and $c_{i}\in EF(v_{i}),i=1,2$,
$d_{S}(EF(v_{1}[c_{1}]),EF(v_{2}[c_{2}]))\geq d_{H}(v_{1},v_{2})$.
If $d_{H}(v_{1},v_{2})=0$, then
$d_{S}(EF(v_{1}[c_{1}]),EF(v_{2}[c_{2}]))=2d_{R}(c_{1},c_{2})$.
\end{lemma}

\begin{cor}
$\C \in \mathcal{G}_{q}(n,k)$ and $d_{S} ( \C )=2\delta$.
\end{cor}

\begin{theorem}
\label{thm:rank_bound} Let $C_v\subseteq \F_q^{m \times t}$ be a
rank-metric code with $d_{R} (C_v)=\delta$, such that all its
codewords are in $EF(v)$ for some binary vector $v$. Let $S$ be
the sub-matrix of $EF(v)$ which corresponds to the dots part of
$EF(v)$. Then the dimension of $C_{v}$ is upper bounded by the
minimum between the number of dots in the last $m-\delta+1$ rows
of $S$ and the number of dots in the first $t-\delta+1$ columns of
$S$.
\end{theorem}

Constructions for codes which attain the bound of
Theorem~\ref{thm:rank_bound} for most important cases are given
in~\cite{EtSi}. Examples are given in the following table
(see~\cite{Sil} for details):

\begin{tabular}{|c|c|c|c|c|}
\hline $q$ & $n$ & $k$ & $d_{s}$ & $| \C |$ \tabularnewline \hline
\hline 2 & 6 & 3 & 4 & 71\tabularnewline \hline 2 & 7 & 3 & 4 &
289\tabularnewline \hline 2 & 8 & 4 & 4 & 4573\tabularnewline
\hline
\end{tabular}

\itwsection{Error-Correcting Projective Space Codes}

Let $\C \in \mathcal{G}_{q}(n,k)$ with $d_{S} (\C)=2\delta$. Let
$Q$ be an $(n-1)$-dimensional subspace of $\F_q^n$ and $v \in
\F_q^n$ such that $v \notin Q$. Let $\C'=\C_1\cup \C_v$, where
$\C_1=\left\{ c\in \C:\: c\subseteq Q\right\} $ and $\C_v=\left\{
c \cap Q:\: c \in \C,~v\in c \right\}$.

\begin{lemma}
$\C' \in \mathcal{P}_{q}(n-1)$ and $d_{S}(\C')=2\delta-1$.
\end{lemma}

By applying this {\it puncturing} method with the 7-dimensional
subspace $Q$ whose generator matrix is\[
\left(\begin{array}{cccccc}
1 & 0 & \ldots & 0 & 0\\
0 & 1 & \ldots & 0 & 0\\
\vdots & \vdots & \ddots & \vdots & \vdots\\
0 & 0 & \ldots & 1 & 0\end{array}\right)\] and the vector
$v=10000001$, on the code with size 4573, and minimum distance 4, in
$\mathcal{G}_{2}(8,4)$, we were able to obtain a code with minimum
distance 3 and size 573 in $\mathcal{P}_{2}(7)$.

\end{itwpaper}

\end{document}